\documentclass[aps,pra,notitlepage,twocolumn,10pt,a4paper]{revtex4-1}

\usepackage{xcolor,graphicx,ulem,soul}
\usepackage{amsmath,amssymb}
\usepackage[colorlinks=true,urlcolor=blue,citecolor=blue,linkcolor=blue]{hyperref}

\begin{document}

\title{Full statistics of homodyne correlation measurements}

\author{B. K\"uhn}\email{benjamin.kuehn2@uni-rostock.de}
\author{W. Vogel}
\affiliation{Arbeitsgruppe Quantenoptik, Institut f\"ur Physik, Universit\"at Rostock, 18051 Rostock, Germany}

\date{\today}

\begin{abstract}
	We derive the full statistics of the product events in homodyne correlation measurements, involving a single mode signal, a local oscillator, a linear optical network, and two linear photodetectors.
	This is performed for the regime of high intensities impinging on the detectors.
	Our description incorporates earlier proposed homodyne correlation measurement schemes, such as the homodyne cross-correlation and homodyne intensity-correlation measurements.
	This analysis extends the amount of information retrieved from such types of measurements, since previously attention was paid only to the expectation value of the correlation statistics.
	As an example, we consider the correlation statistics of coherent, Gaussian, and Fock states.
	Moreover, nonclassical light is certified on the basis of the variance of the measurement outcome.
\end{abstract}

\maketitle


\section{Introduction}
	
	In the past few decades a diversity of different detection schemes has been proposed to gain information about the quantum state of light fields.
	A prominent example is balanced homodyning \cite{Yuen1983,Welsch1999,Lvovsky2009}, which allows one to obtain the field strength statistics for different optical phases \cite{Grabow1993}.
	By contrast, the unbalanced homodyne detection \cite{Wallentowitz1996} gives access to the photon-number distribution of the coherently displaced signal.
	The measurement outcomes of both schemes provide the full information on the quantum state. 
	Hence they can be transformed to other state representations, such as quasiprobabilities \cite{Wigner1932,Husimi1940,Cahill1969,Smithey1993,Vogel1989,Smithey1993,Kiesel2008} or the density matrix \cite{Smithey1993,Kuehn1994,Leonhardt1996,Mancini1997}.
	Furthermore, balanced eight-port homodyning \cite{Noh1991,Noh1993} allows one to directly measure the Husimi $Q$~function \cite{Freyberger1992}.	
	
	Small quantum efficiencies significantly smooth out the nonclassical effects.
	For such conditions, homodyne correlation measurement (HCM) techniques have been developed \cite{Vogel1991,Vogel1995}, where the quantum efficiency merely rescales the measurement outcome, due to the detection of normal-ordered quantities.
	Similar to balanced homodyne detection, these experimental setups rely on the interference of a signal beam with coherent light on beam splitters and the intensity detection of two outgoing beams.
	Instead of analyzing the difference signal of the two photodetectors, the correlated fluctuations of the photoelectric currents are studied. 
	Later on, another method was proposed and applied in experiments, which is based on balanced homodyne detection conditioned on a photon-number measurement \cite{Carmichael2000,Reiner2001,Gerber2009}. 
	It yields similar insight in the quantum properties of light as HCMs. 
	
	In Ref. \cite{Vogel1995}, two different realizations of HCMs were studied, the homodyne intensity-correlation measurement and the homodyne cross-correlation measurement.	
	Recently, both techniques have been successfully implemented in experiments.
	In particular, the homodyne intensity-correlation measurement was implemented by following the original proposal in Ref. \cite{Vogel1991} to certify quadrature squeezing in resonance fluorescence light from a single quantum dot \cite{Schulte2015}. 
	The homodyne cross-correlation measurement, on the other hand, demonstrated the existence of anomalous quantum correlations of field strength and intensity noise of squeezed light \cite{Kuehn2017}, which even extends beyond the phase interval of squeezing.
	As an extension, multiport schemes have been considered, which give access to higher-order normal-ordered moments of the phase-dependent quadrature operator \cite{Shchukin2006} and the displaced photon-number operator \cite{Kuehn2016}.
	
	Until now, only the mean of the product of the fluctuations of the photoelectric currents of the detectors in such schemes was considered. 
	However, the exact shape of the full product statistics is yet unknown.
	In this work, we close this gap by deriving a closed expression for the full HCM statistics and we also determine the associated positive-operator-valued measure (POVM).
	As an application, we develop a nonclassicality criterion based on the variance of these statistics and demonstrate its usefulness to certify the nonclassicality of an amplitude-squeezed coherent state.
		
	Our work is organized as follows.
	In Sec. \ref{ch:correlationmeasurement} we recall earlier proposed homodyne correlation measurement schemes and consider them as specific configurations of a more general measurement device including a linear optical network and two linear standard detectors.
	The full correlation statistics of the product of the photocurrent fluctuations is derived in Sec. \ref{ch:JointProb}.
	In Sec. \ref{ch:examples} we study the correlation statistics of several states, such as coherent states, Gaussian states, and Fock states.
	Furthermore, we relate our result to the certification of anomalous quantum correlations in Sec. \ref{ch:nonclassicality} and we provide a sufficient nonclassicality condition based on the detection outcome of our correlation measurement device.
	We summarize in Sec. \ref{ch:Conclusions}.
	
\section{Correlation measurement with two linear detectors}
\label{ch:correlationmeasurement}

	\begin{figure}[h]
		\includegraphics[clip,scale=0.47]{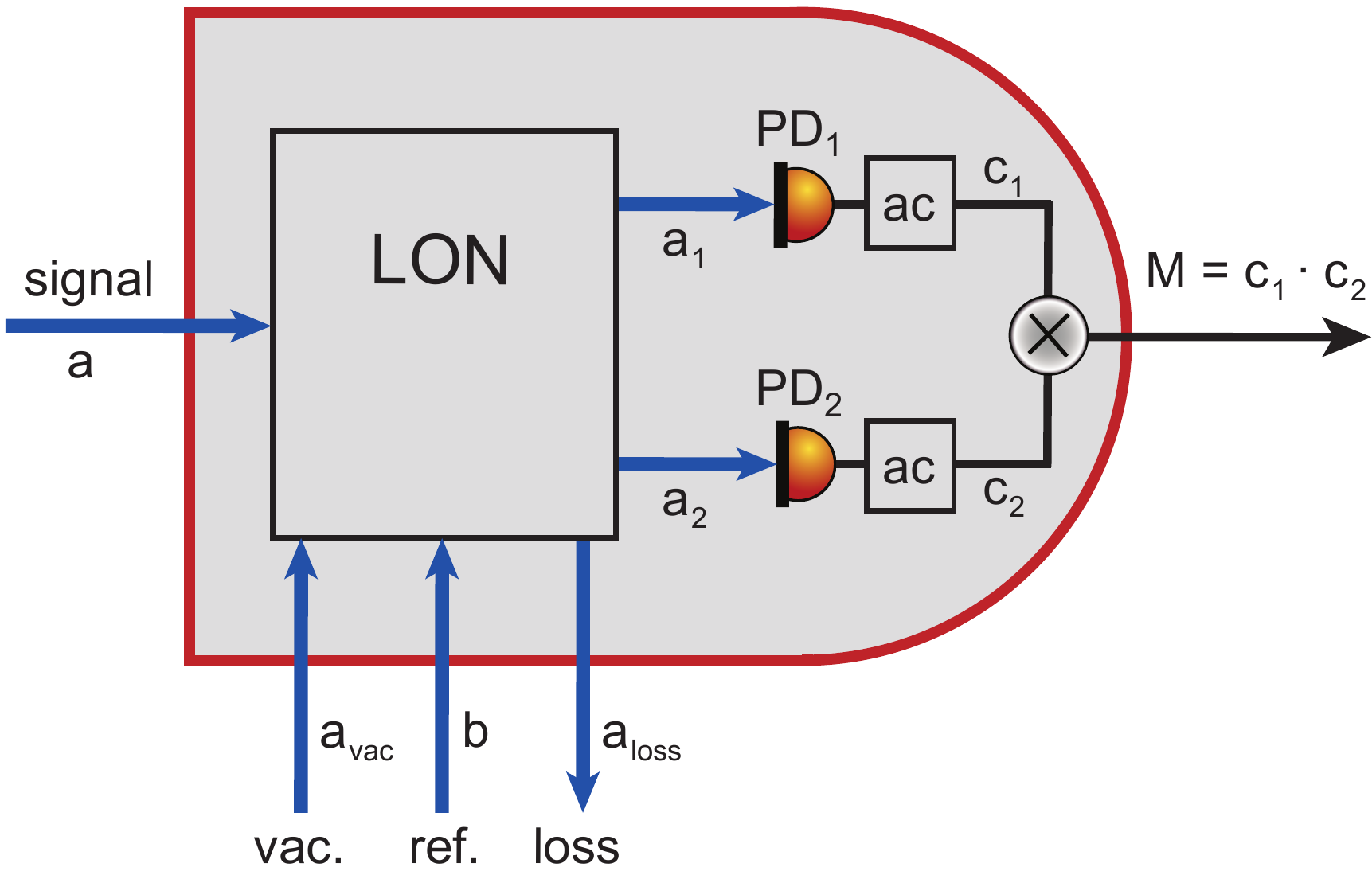}
		\caption{
		The HCM detector device for quantum light (gray shaded area).
		A linear optical network (LON) combines the signal field $\hat a$ with a reference mode $\hat b$, and a vacuum input $\hat a_{\mathrm{vac}}$.
		Two output beams $\hat a_1$ and $\hat a_2$ are detected by linear photodetectors, $\mathrm{PD}_1$ and $\mathrm{PD}_2$, with efficiencies $\eta_1$ and $\eta_2$ and dark noise counts $\nu_1$ and $\nu_2$.
		They record the intensity noise of both output modes through the alternating photoelectric currents (ac) $c_1$ and $c_2$.
		The two currents are multiplied, resulting in the measurement outcome $M$.
		Possible losses are incorporated in the loss mode $\hat a_{\mathrm{loss}}$.
		}\label{fig:Setup}
	\end{figure}

	Consider the scheme, which is illustrated in Fig. \ref{fig:Setup}.
	A signal beam, a reference beam, and an additional vacuum input, characterized by the photon annihilation operators $\hat a$, $\hat b$, and $\hat a_{\mathrm{vac}}$, respectively, are combined by a linear optical network (LON).
	The latter is usually implemented by an assembly of passive linear optical elements, such as beam splitters, which realize a unitary transformation of the three input modes. 
	The LON outputs are three beams, in particular, $\hat a_1$, $\hat a_2$, and a loss mode $\hat a_{\mathrm{loss}}$.
	Constant loss of the involved beam splitters in the LON is transferred to the output $\hat a_{\mathrm{loss}}$.
	The photonic operators of the outgoing fields $\hat a_1$ and $\hat a_2$ are related to that ones of the input fields by the linear input-output relation	
	\begin{align}\label{eq:inout}
		\begin{pmatrix}
			\hat a_{1}\\
			\hat a_{2}\\
		\end{pmatrix}
		&=\mathcal{Q}
		\begin{pmatrix}
			\hat a\\
			\hat b\\
			\hat a_{\mathrm{vac}}\\
		\end{pmatrix}
		, 	
	\end{align}
	where the $2\times 3$ matrix $\mathcal{Q}$ is a submatrix of a unitary $3\times 3$ matrix.
	In particular, $\mathcal{Q}$ excludes the output $\hat a_{\mathrm{loss}}$, incorporating losses.
	It only keeps the modes $\hat a_1$ and $\hat a_2$ that are relevant in the following considerations.
	To be more specific, we consider in the further calculations only expectation values over functions of $\hat a_1$ and $\hat a_2$. 
	Therefore, the loss mode is simply traced out.
	
	The intensity correlations of the beams $\hat a_1$ and $\hat a_2$ are recorded by linear photodetectors with efficiencies $\eta_1$ and $\eta_2$ and dark noise counts $\nu_1$ and $\nu_2$, respectively.
	The photoelectric current fluctuations $c_1$ and $c_2$ of the two detectors are extracted by applying electronic filters and they are multiplied afterward.
	The outcome, $M=c_1\cdot c_2$, contains information about the intensity noise correlation of the two modes $\hat a_1$ and $\hat a_2$ and consequently also about the signal field.
	
	Two types of such HCM devices have been studied. 
	The first one is the homodyne intensity-correlation measurement, which was introduced in Ref. \cite{Vogel1991} and analyzed in more detail in \cite{Vogel1995}.
	It employs two beam splitters.
	In the first step, the signal field interferes with a coherent local oscillator $|\alpha_\mathrm{L}\rangle$ (in the reference channel) on the first beam splitter with field transmittance $T_1$ and reflectance $R_1$.
	One of the outputs is then split (combined with vacuum) by the second beam splitter of field transmittance $T_2$ and reflectance $R_2$.
	The two outgoing beams of this second beam splitter correspond to the two modes $\hat a_1$ and $\hat a_2$ in Fig. \ref{fig:Setup}.
	For the associated input-output matrix, we obtain 
	\begin{align}\label{eq:ic}
		\mathcal{Q}^{(\mathrm{ic})}&=
		\begin{pmatrix}
			T_2T_1 & T_2R_1 & R_2\\
			R_2T_1 & R_2R_1 & T_2\\
		\end{pmatrix}.
	\end{align}	
	This measurement technique was recently applied for  the detection of quadrature squeezing in the resonance fluorescence of a two-level system \cite{Schulte2015}.
	For the presence of constant losses in the LON, the input-output matrix has to be adjusted as $|T_j|^2+|R_j|^2<1$ for $j=1,2$.
	
	The second type of HCM device, referred to as the homodyne cross-correlation scheme, was introduced in Ref. \cite{Vogel1995}. 
	In this four-port optical setup the signal beam is combined on a beam splitter (field strength transmittance $T$ and reflectance $R$) with a local oscillator (LO), which is prepared in a coherent state $|\alpha_\mathrm{L}\rangle$. 
	The two output beams are the modes $\hat a_1$ and $\hat a_2$ in Fig. \ref{fig:Setup}.
	In the absence of losses the input-output matrix in Eq. \eqref{eq:inout} reads in this case
	\begin{align}\label{eq:cc}
		\mathcal{Q}^{(\mathrm{cc})}&=
		\begin{pmatrix}
			T & R & 0\\
			R & T & 0\\
		\end{pmatrix},
	\end{align}
	with $|T|^2+|R|^2=1$ and $T^\ast R+R^\ast T=0$.
	Possible losses would result in nonzero elements in the third column of this matrix.
	Recently, it was demonstrated experimentally that such a device can detect an anomalous quantum correlation of two noncommuting observables for a phase-squeezed coherent state \cite{Kuehn2017}.
	
\section{Full correlation statistics}
\label{ch:JointProb}	
	
	In the earlier works, which considered a correlation detector of the kind described in the preceding section, only the expectation value of the measurement outcome $\mathbb{E}\left(M\right)=\mathbb{E}\left(c_1\cdot c_2\right)$ was considered to extract information about the signal field.
	In the present paper, however, we investigate the full correlation statistics, which includes the higher-order moments of $M$.  
	For this purpose we will determine the full correlation statistics $w(M)$ in this section.	
			
	Applying the photon-counting theory of two detectors with efficiencies $\eta_1$ and $\eta_2$ and independent dark noise counts $\nu_1$ and $\nu_2$, the joint probability that detector $\mathrm{PD}_1$ records $m_1$ events and detector $\mathrm{PD}_2$ records $m_2$ events is given by \cite{Kelley1964,Glauber1965}
	\begin{align}\label{eq:prodstat1}
	\begin{aligned}
		\mathcal{P}_{m_1,m_2}&=\left\langle:\dfrac{\left(\eta_1\hat n_1+\nu_1\right)^{m_1}}{m_1!}\,\mathrm e^{-(\eta_1\hat n_1+\nu_1)}\right.\\
		&\left.\times\dfrac{\left(\eta_2\hat n_2+\nu_2\right)^{m_2}}{m_2!}\,\mathrm e^{-(\eta_2\hat n_2+\nu_2)}:\right\rangle.
	\end{aligned}
	\end{align}
	Here $:\cdot:$ denotes normal ordering, $\langle\cdot\rangle$ is the quantum mechanical expectation value, and $\hat n_j=\hat a_j^\dagger\hat a_j$ are the photon-number operators of the output fields $\hat a_1$ and $\hat a_2$.
	They are related to the input beams $\hat a$, $\hat b$, and $\hat a_{\mathrm{vac}}$ of the LON by Eq. \eqref{eq:inout}.
	In the following we assume that the reference mode $\hat b$ is prepared in a coherent state $|\alpha_\mathrm{L}\rangle$ ($\alpha_{\mathrm{L}}=|\alpha_{\mathrm{L}}|e^{i\phi}$), as is the case for both the homodyne intensity-correlation and the homodyne cross-correlation measurements \cite{Vogel1995}.
	Additionally, we demand that the intensities of the input modes $\hat a$ and $\hat b$ result in bright light in the LON outputs, i.e., $\eta_1\langle\hat n_1\rangle\gg 1$ and $\eta_2\langle\hat n_2\rangle\gg 1$, such that photon-number resolution of the two detectors is not required. 
	
	The quantum state $\hat\rho$ of the signal can be represented in terms of coherent states $|\alpha\rangle$ by means of the Glauber-Sudarshan $P$~function as \cite{Glauber1963,Sudarshan1963}
	\begin{align}\label{eq:signal}
		\hat\rho&=\int d^2\alpha\,P(\alpha)|\alpha\rangle\langle\alpha|.
	\end{align}
	Using $\hat a_{\mathrm{vac}}|0\rangle=0$, $\hat b|\alpha_{\mathrm{L}}\rangle=\alpha_{\mathrm{L}}|\alpha_{\mathrm{L}}\rangle$, and the input-output relation \eqref{eq:inout}, we can rewrite Eq. \eqref{eq:prodstat1} as
	\begin{align}\label{eq:Poisson}
	\begin{aligned}
		\mathcal{P}_{m_1,m_2}&=\int d^2\alpha\,P(\alpha)\\
		&\times\dfrac{\left(\eta_1|\alpha_1(\alpha,\alpha_\mathrm{L})|^2+\nu_1\right)^{m_1}}{m_1!}\,\mathrm e^{-(\eta_1|\alpha_1(\alpha,\alpha_\mathrm{L})|^2+\nu_1)}\\
		&\times\dfrac{\left(\eta_2|\alpha_2(\alpha,\alpha_\mathrm{L})|^2+\nu_2\right)^{m_2}}{m_2!}\,\mathrm e^{-(\eta_2|\alpha_2(\alpha,\alpha_\mathrm{L})|^2+\nu_2)},
	\end{aligned}
	\end{align}
	where 
	\begin{align}\label{eq:alphak}
		\alpha_j(\alpha,\alpha_\mathrm{L})&=\mathcal{Q}_{j1}\alpha+\mathcal{Q}_{j2}\alpha_{\mathrm{L}},\,\,\,\,\,\,\,\,j=1,2,
	\end{align}
	are the coherent amplitudes of the two outgoing fields $\hat a_1$ and $\hat a_2$ in the case of a coherent signal with amplitude $\alpha$ and $\mathcal{Q}_{ju}$ is the element of the input-output matrix $\mathcal{Q}$ in row $j=1,2$ and column $u=1,2,3$.
	In the limit $\eta_1|\alpha_1(\alpha,\alpha_\mathrm{L})|^2\gg 1$ and $\eta_2|\alpha_2(\alpha,\alpha_\mathrm{L})|^2\gg 1$, one can replace the Poisson distributions in the integrand of Eq. \eqref{eq:Poisson} by Gaussian distributions, where the discrete number events $(m_1,m_2)$ are replaced by continuous variables $(x_1,x_2)$. 
	In this regime the joint event statistics reads
	\begin{align}\label{eq:joint}
	\begin{aligned}
		\mathcal{P}(x_1,x_2)&=\int d^2\alpha\,P(\alpha)\dfrac1{\sqrt{2\pi[\eta_1|\alpha_1(\alpha,\alpha_\mathrm{L})|^2+\nu_1]}}\\
		&\times\dfrac1{\sqrt{2\pi[\eta_2|\alpha_2(\alpha,\alpha_\mathrm{L})|^2+\nu_2]}}\\
		&\times\exp\left(-\dfrac{[x_1-\eta_1|\alpha_1(\alpha,\alpha_\mathrm{L})|^2-\nu_1]^2}{2[\eta_1|\alpha_1(\alpha,\alpha_\mathrm{L})|^2+\nu_1]}\right)\\
		&\times\exp\left(-\dfrac{[x_2-\eta_2|\alpha_2(\alpha,\alpha_\mathrm{L})|^2-\nu_2]^2}{2[\eta_2|\alpha_2(\alpha,\alpha_\mathrm{L})|^2+\nu_2]}\right).
	\end{aligned}
	\end{align}
	This result was already derived in Ref. \cite{Grabow1993}, but instead of calculating the difference statistics, in the following we determine the product statistics of the photoelectric current fluctuations $c_j=x_j-\langle x_j\rangle$ ($j=1,2$).
	Here the mean photoelectric current of detector $\mathrm{PD}_j$ is readily derived as
	\begin{align}\label{eq:meanx}
		\langle x_j\rangle&=\eta_j\int d^2\alpha\,P(\alpha)\,|\alpha_j(\alpha,\alpha_\mathrm{L})|^2+\nu_j.
	\end{align}
	In a first step we determine the joint statistics $p(c_1,c_2)$, which is obtained out of Eq. \eqref{eq:joint} by the relation
	\begin{align}\label{eq:prodfluctpre}
		p(c_1,c_2)&=\mathcal{P}(\langle x_1\rangle+c_1,\langle x_2\rangle+c_2).
	\end{align}	
	Therefore, it can be expressed in terms of the $P$~function of the signal as
	\begin{align}\label{eq:prodfluct}
	\begin{aligned}
		p(c_1,c_2)&=\int d^2\alpha\,P(\alpha)\,\dfrac1{\sqrt{2\pi\sigma_1^2(\alpha)}}\exp\left(-\dfrac{[c_1-\mu_1(\alpha)]^2}{2\sigma_1^2(\alpha)}\right)\\
		&\times \dfrac1{\sqrt{2\pi\sigma_2^2(\alpha)}}\exp\left(-\dfrac{[c_2-\mu_2(\alpha)]^2}{2\sigma_2^2(\alpha)}\right),
	\end{aligned}
	\end{align}
	where we defined the variances
	\begin{align}\label{eq:sigmapre}
		\sigma_j^2(\alpha)&=\eta_j|\alpha_j(\alpha,\alpha_\mathrm{L})|^2+\nu_j,\,\,\,\,\,\,\,\,j=1,2,
	\end{align}
	and the means	
	\begin{align}\label{eq:meanpre}
		\mu_j(\alpha)&=\eta_j|\alpha_j(\alpha,\alpha_\mathrm{L})|^2+\nu_j-\langle x_j\rangle,\,\,\,\,\,\,j=1,2.
	\end{align}
	Let us rewrite Eq. \eqref{eq:alphak} as
	\begin{align}\label{eq:decompositionnoise}
	\begin{aligned}
		\alpha_j(\alpha,\alpha_\mathrm{L})&=\mathcal{Q}_{j1}\gamma(\alpha)+\alpha_j(\langle\hat a\rangle,\alpha_\mathrm{L}),\,\,\,\,\,\,j=1,2,
	\end{aligned}
	\end{align}
	where $\langle\hat a\rangle=\int d^2\alpha\,P(\alpha)\alpha$ is the mean signal amplitude and $\gamma(\alpha)=\alpha-\langle\hat a\rangle$ is the signal noise.
	Since we consider the limit $\eta_j|\alpha_j(\alpha,\alpha_\mathrm{L})|^2\gg 1$, the signal noise $\gamma(\alpha)$ in the decomposition \eqref{eq:decompositionnoise} is small compared to the mean interference amplitude $\alpha_j(\langle\hat a\rangle,\alpha_\mathrm{L})$.
	Accordingly, the variances in Eq. \eqref{eq:sigmapre} are in this approximation independent of the signal fluctuations $\gamma(\alpha)$ and depend only on the mean signal amplitude $\langle\hat a\rangle$, i.e.,
	\begin{align}\label{eq:sigma}
		\sigma^2_j=\eta_j\left|\alpha_j(\langle\hat a\rangle,\alpha_\mathrm{L})\right|^2+\nu_j,\,\,\,\,\,\,j=1,2.
	\end{align}
	Inserting Eq. \eqref{eq:meanx} into Eq. \eqref{eq:meanpre} and considering the same approximation, one derives that the means in Eq. \eqref{eq:meanpre} reduce to a function only of the noise $\gamma=\gamma(\alpha)$, i.e.,
	\begin{align}\label{eq:mean}
		\mu_j(\gamma)=h_j^\ast\gamma+h_j\gamma^\ast,\,\,\,\,\,\,j=1,2,
	\end{align}	
	where
	\begin{align}\label{eq:h}
		h_j=\eta_j\mathcal{Q}^\ast_{j1}\alpha_j(\langle\hat a\rangle,\alpha_\mathrm{L}),\,\,\,\,\,\,j=1,2.
	\end{align}
	Now we use Eqs. \eqref{eq:sigma} and \eqref{eq:mean} in the joint statistics of the photoelectric current fluctuations in Eq. \eqref{eq:prodfluct} and we additionally substitute the integration variable $\alpha$ for the noise amplitude $\gamma=\alpha-\langle\hat a\rangle$.
	This yields the result
	\begin{align}\label{eq:jointfluct}
	\begin{aligned}
		p(c_1,c_2)&=\int d^2\gamma\,P(\gamma+\langle\hat a\rangle)\\
		&\times\dfrac1{\sqrt{2\pi\sigma_1^2}}\exp\left(-\dfrac{[c_1-\mu_1(\gamma)]^2}{2\sigma_1^2}\right)\\
		&\times\dfrac1{\sqrt{2\pi\sigma_2^2}}\exp\left(-\dfrac{[c_2-\mu_2(\gamma)]^2}{2\sigma_2^2}\right).
	\end{aligned}
	\end{align}
	Finally, one can determine the probability distribution $w(M)$ of the product $M=c_1\cdot c_2$ of the photoelectric current fluctuations on the basis of the joint statistics in Eq. \eqref{eq:jointfluct} by using the relation
	\begin{align}\label{eq:howcalculatewm}
		w(M)=\int_{-\infty}^\infty \dfrac{dy}{|y|}\,p(y,M/y).
	\end{align}
	In Ref. \cite{Cui2016} the exact probability distribution of the product of two real Gaussian random variables with nonzero means was derived by utilizing Eq. \eqref{eq:howcalculatewm}.
	Since the joint statistics $p(c_1,c_2)$ in Eq. \eqref{eq:jointfluct} is a combination (weighted with the signal $P$ function) of Gaussian probability distributions of uncorrelated random variables $c_1$ and $c_2$, one can directly apply the results of this reference.
	Therefore, we obtain the correlation statistics
	\begin{align}\label{eq:fullwm}
	\begin{aligned}
		w(M)&=\dfrac1{\sigma_1\sigma_2}\int d^2\gamma\, P(\gamma+\langle\hat a\rangle)\\
		&\times\sum_{u=0}^\infty\sum_{\ell=0}^{2u}\mathcal{W}_{u,\ell}\left(\dfrac{M}{\sigma_1\sigma_2}\right)\mathcal{G}_{\ell,2u-\ell}(\gamma,\gamma^\ast)
	\end{aligned}
	\end{align}
	for an arbitrary signal state $\hat\rho$ in Eq. \eqref{eq:signal}.
	Here we introduced the functions
	\begin{align}
	\begin{aligned}
		\mathcal{G}_{a,b}(\gamma,\gamma^\ast)&=\left[\dfrac{\mu_1(\gamma)}{\sigma_1}\right]^a\left[\dfrac{\mu_2(\gamma)}{\sigma_2}\right]^b\exp\left[-\dfrac{\mu_1^2(\gamma)}{2\sigma^2_1}-\dfrac{\mu_2^2(\gamma)}{2\sigma_2^2}\right],\\
		\mathcal{W}_{a,b}(z)&=\dfrac1{\pi}\,\dfrac1{(2a)!}\binom{2a}{b}z^{2a-b}|z|^{b-a}K_{b-a}(|z|)
	\end{aligned}
	\end{align}
	with the modified Bessel functions of the second kind $K_v(\cdot)$.	
	Note that general nonclassical states, for which the $P$~function is not a classical probability density \cite{Titulaer1965}, are included in this expression.
	
	Let us introduce the operator for the signal amplitude fluctuation as
	\begin{align}
		\delta \hat a=\hat a-\langle\hat a\rangle,
	\end{align}
	which fulfills the bosonic commutation relations.
	We find the POVM for the measurement outcome $M$ of our HCM device as the normal-ordered operator
	\begin{align}\label{eq:POVM}
		\hat \Pi_M=\,:\dfrac1{\sigma_1\sigma_2}\sum_{u=0}^\infty\sum_{\ell=0}^{2u}\mathcal{W}_{u,\ell}\left(\dfrac{M}{\sigma_1\sigma_2}\right)\hat{\mathcal{G}}_{\ell,2u-\ell}(\delta\hat a,\delta \hat a^\dagger):\,.
	\end{align}
	It holds that $\int dM\,\hat \Pi_M=\hat 1$, since 
	\begin{align}
		w(M)=\langle\hat \Pi_M\rangle
	\end{align}
	yields the correlation statistics.
	The POVM together with the full correlation statistics in Eq. \eqref{eq:fullwm} is a central result of this work.
		
	Now we determine the expectation value $\mathbb{E}\left(M\right)$ and the variance $\mathrm{var}\left(M\right)$ of $M$, since it is needed for consideration in the following sections.
	For this purpose, we first calculate these quantities for two uncorrelated Gaussian random variables $c_1$ and $c_2$ with variances $\sigma_1^2$ and $\sigma_2^2$ [cf. Eq. \eqref{eq:sigma}] and means $\mu_1(\gamma)$ and $\mu_2(\gamma)$ [cf. Eq. \eqref{eq:mean}], respectively, conditioned on the value of $\gamma$.
	By using the well-known results for the moments of Gaussian distributed variables (see, e.g., Ref. \cite{Gradshteyn2000}), they are given by
	\begin{align}
	\begin{aligned}
		\mathbb{E}\left(M^k|\gamma\right)&=\mathbb{E}\left(c_1^k\cdot c_2^k|\gamma\right)\\
		&=\left(-\dfrac{\sigma_1\sigma_2}{2}\right)^k\,H_k\left(i\dfrac{\mu_1(\gamma)}{\sqrt{2}\sigma_1}\right)H_k\left(i\dfrac{\mu_2(\gamma)}{\sqrt{2}\sigma_2}\right),
	\end{aligned}
	\end{align}
	where $H_n(\cdot)$ are the Hermite polynomials.
	In particular, the first and second conditional moments read
	\begin{align}\label{eq:expcond}
		\mathbb{E}\left(M|\gamma\right)&=\mu_1(\gamma)\,\mu_2(\gamma)
	\end{align}
	and
	\begin{align}\label{eq:secondcond}
		\mathbb{E}\left(M^2|\gamma\right)&=\left[\mu_1^2(\gamma)+\sigma_1^2\right]\,\left[\mu_2^2(\gamma)+\sigma_2^2\right].
	\end{align}
	Combining this with Eq. \eqref{eq:jointfluct}, yields the expectation value and the variance of $M$ for the signal state $\hat\rho$ under consideration as
	\begin{align}\label{eq:Mexp}
		\mathbb{E}\left(M\right)=\int d^2\gamma\,P(\gamma+\langle\hat a\rangle)\,\mu_1(\gamma)\mu_2(\gamma)
	\end{align}
	and
	\begin{align}\label{eq:Mvar}
	\begin{aligned}
		\mathrm{var}\left(M\right)&=\mathbb{E}\left(M^2\right)-\mathbb{E}\left(M\right)^2\\
		&=\int d^2\gamma\,P(\gamma+\langle\hat a\rangle)\,\left[\mu_1^2(\gamma)+\sigma_1^2\right]\,\left[\mu_2^2(\gamma)+\sigma_2^2\right]\\
		&-\left(\int d^2\gamma\,P(\gamma+\langle\hat a\rangle)\,\mu_1(\gamma)\,\mu_2(\gamma)\right)^2.
	\end{aligned}
	\end{align}
	We want to point out that the expectation value in Eq. \eqref{eq:Mexp} does not depend on the independent dark noise counts $\nu_1$ and $\nu_2$ of the two detectors in Fig. \ref{fig:Setup}.
	This can be easily seen from Eq. \eqref{eq:mean} together with Eq. \eqref{eq:h}, where $\mu_1(\gamma)$ and $\mu_2(\gamma)$ are independent of the dark noise.
	By contrast, this kind of noise may contaminate the higher moments, such as the variance in Eq. \eqref{eq:Mvar}.
	Note that the latter incorporates the quantities $\sigma_1^2$ and $\sigma_2^2$, which depend on the dark noise counts $\nu_1$ and $\nu_2$ [cf. Eq. \eqref{eq:sigma}].
		
\section{Examples of correlation statistics}
\label{ch:examples}

	In this section we determine the correlation statistics for different states of the signal input in Fig. \ref{fig:Setup}.
	In particular, we study arbitrary Gaussian states, which include the coherent states. 
	In addition, we also consider Fock states.
	
\subsection{Coherent states}
\label{sub:coherent}
	
	Let us start with the simplest case of coherent states $|\alpha\rangle$, which have the mean amplitude $\langle\hat a\rangle=\alpha$ and are represented by the $P$~function $P(\chi)=\delta(\chi-\alpha)$, with $\delta(\cdot)$ being the Dirac delta function.
	The corresponding correlation statistics is readily derived from Eq. \eqref{eq:POVM} and reads
	\begin{align}\label{eq:wMcoh}
		w(M)=\langle\alpha|\hat \Pi_M|\alpha\rangle=\dfrac1{\pi\sigma_1\sigma_2}K_0\left(\dfrac{|M|}{\sigma_1\sigma_2}\right).
	\end{align}
	Note that the amplitude $\alpha$ still enters through the variances $\sigma_j^2$ [cf. Eq. \eqref{eq:sigma}] and influences the widths of this probability distribution.
	
	\begin{figure}[h]
	\centering
 		\includegraphics[width=0.99\linewidth]{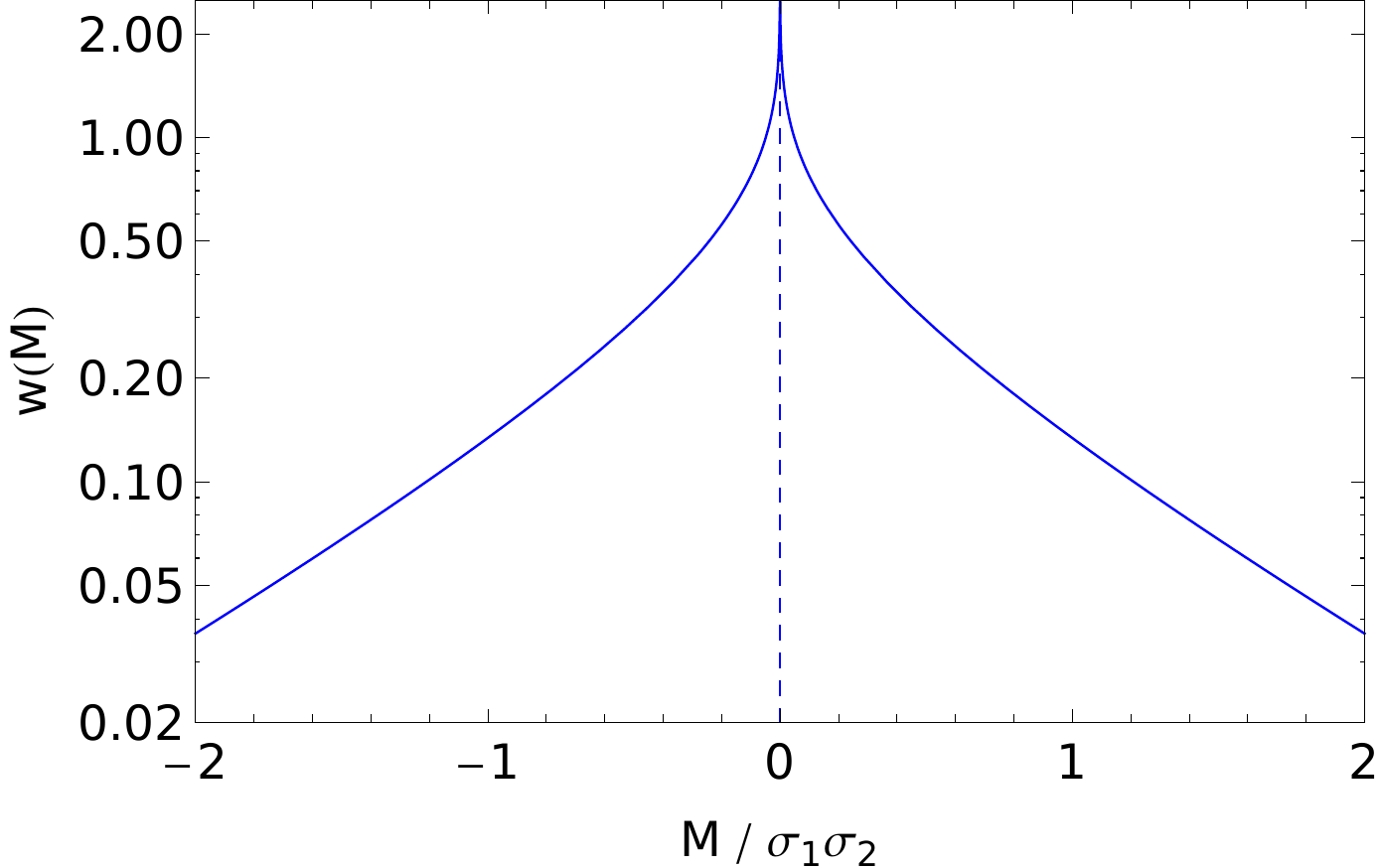}
		\caption{
			Correlation statistics $w(M)$ for the signal beam prepared in a coherent state.
			Note that $M$ is normalized to $\sigma_1\sigma_2$ [cf. Eq. \eqref{eq:sigma}].
			The vertical dashed line indicates the expectation value $\mathbb{E}(M)$.
		}
		\label{fig:coh}
	\end{figure}	
	
	The statistics in Eq. \eqref{eq:wMcoh} is shown in Fig. \ref{fig:coh}.
	It is symmetric with respect to $M$ and consequently the expectation value $\mathbb{E}\left(M\right)$ is zero [see also Eq. \eqref{eq:Mexp} together with Eq. \eqref{eq:mean}].	
	This is reasonable as the LON in Fig. \ref{fig:Setup} leads for coherent input states to coherent output states, which show no intensity noise correlation in its two modes $\hat a_1$ and $\hat a_2$.
	Although the input state is Gaussian, we see that the statistics of the measurement outcome $M$ reveals a strongly non-Gaussian shape, which even has a singularity at $M=0$. 
		
\subsection{Gaussian states}
\label{sub:gaussian}
	
	Next we consider Gaussian states, which are completely described by their first and second moments.
	Alternatively, one can uniquely define them by the maximal and minimal quadrature variance, $V_{\mathrm{x}}$ and $V_{\mathrm{p}}$, respectively, together with the mean amplitude $\langle\hat a\rangle$ and the orientation angle $\phi_\xi$ in phase space.  
	The Glauber-Sudarshan $P$~function of these Gaussian states can be highly singular, while their characteristic function
	\begin{align}\label{eq:characsq}
	\begin{aligned}
		\Phi(\beta)&=\exp\left[\beta\langle\hat a^\dagger\rangle-\beta^*\langle\hat a\rangle\right]\exp\left[-\dfrac{\beta\beta^*}{4}\left(V_{\mathrm{x}}+V_{\mathrm{p}}-2\right)\right]\\
		&\times\exp\left[-\dfrac{\beta^2}{8}e^{-i\phi_\xi}\left(V_{\mathrm{x}}-V_{\mathrm{p}}\right)-\dfrac{\beta^{*2}}{8}e^{i\phi_\xi}\left(V_{\mathrm{x}}-V_{\mathrm{p}}\right)\right]
	\end{aligned}
	\end{align}
	is always a regular function.
	In order to derive the joint statistics of the photoelectric current fluctuations in Eq. \eqref{eq:jointfluct} and on this basis the correlation statistics $w(M)$, it is therefore convenient to express the $P$~function in terms of the characteristic function by
	\begin{align}\label{eq:PPhi}
		P(\gamma)=\dfrac1{\pi^2}\int d^2\beta\,e^{\gamma\beta^\ast-\gamma^\ast\beta}\Phi(\beta).
	\end{align}		
	The definition of the Gaussian states in Eq. \eqref{eq:characsq} includes mixed states for which $V_{\mathrm{x}}V_{\mathrm{p}}>1$.
	The special case $V_{\mathrm{x}}V_{\mathrm{p}}=1$ yields pure Gaussian states with the coherent states obtained for $V_{\mathrm{x}}=V_{\mathrm{p}}=1$. 
	For a minimal quadrature variance $V_{\mathrm{p}}<1$, we refer to the Gaussian state as a squeezed coherent state.
	In the following, we assume, without loss of generality, that the mean amplitude $\langle\hat a\rangle$ is real.
	Furthermore, if $\phi_\xi=0$, the state is referred to as an amplitude-squeezed coherent state and it is called a phase-squeezed coherent state for $\phi_\xi=\pi$.
			
	Using the relation \eqref{eq:PPhi} in Eq. \eqref{eq:jointfluct}, we obtain, after a straightforward calculation involving Gaussian integrals, the joint probability distribution for Gaussian states as
	\begin{align}
	\begin{aligned}
		p(c_1,c_2)&=\dfrac1{2\pi s_1s_2\sqrt{1-\mathcal{C}^2}}\\
		&\times\exp\left\{-\dfrac1{2(1-\mathcal{C}^2)}\left[\dfrac{c_1^2}{s_1^2}-\dfrac{2\mathcal{C} c_1 c_2}{s_1s_2}+\dfrac{c_2^2}{s_2^2}\right]\right\}.
	\end{aligned}
	\end{align}	
	\begin{figure}[h]
	\centering
 		\begin{minipage}[t]{1.0\linewidth} 
			\includegraphics[width=0.85\linewidth]{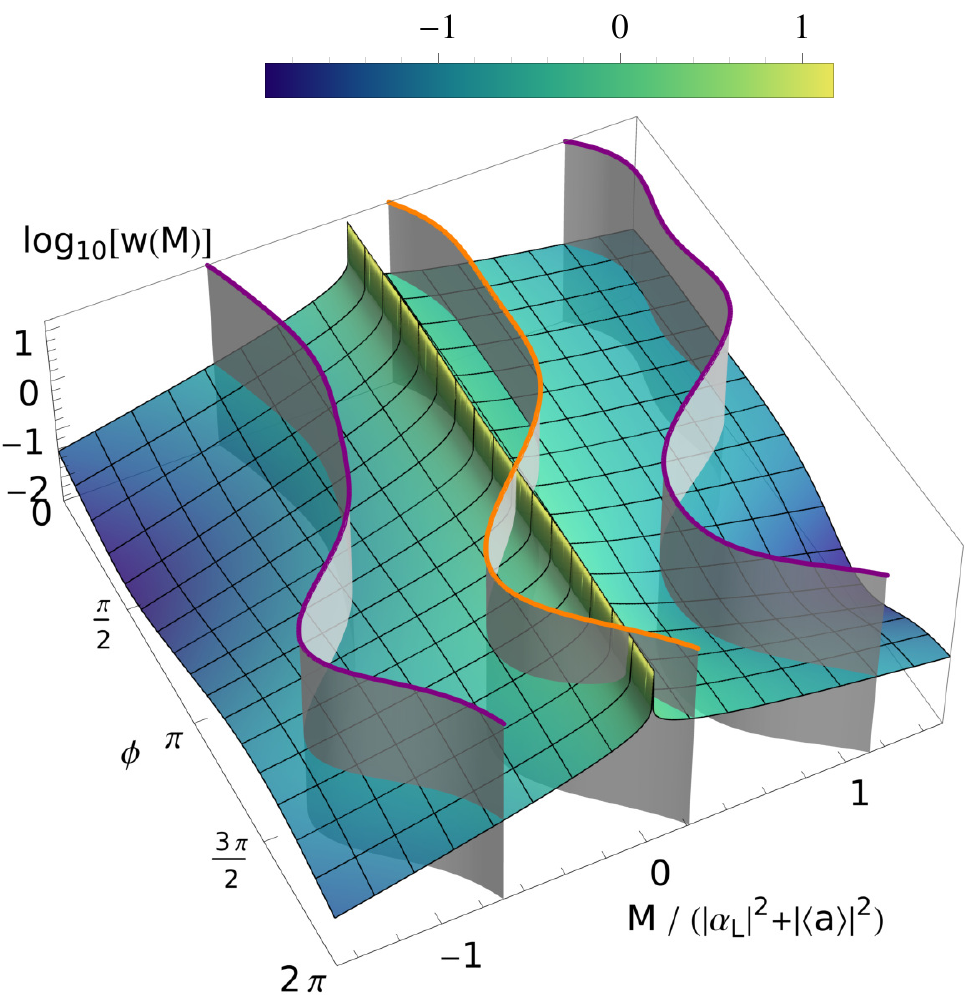}
			\label{fig:gaussPhaseSq}
 			\\\vspace*{3ex}
 			\includegraphics[width=0.85\linewidth]{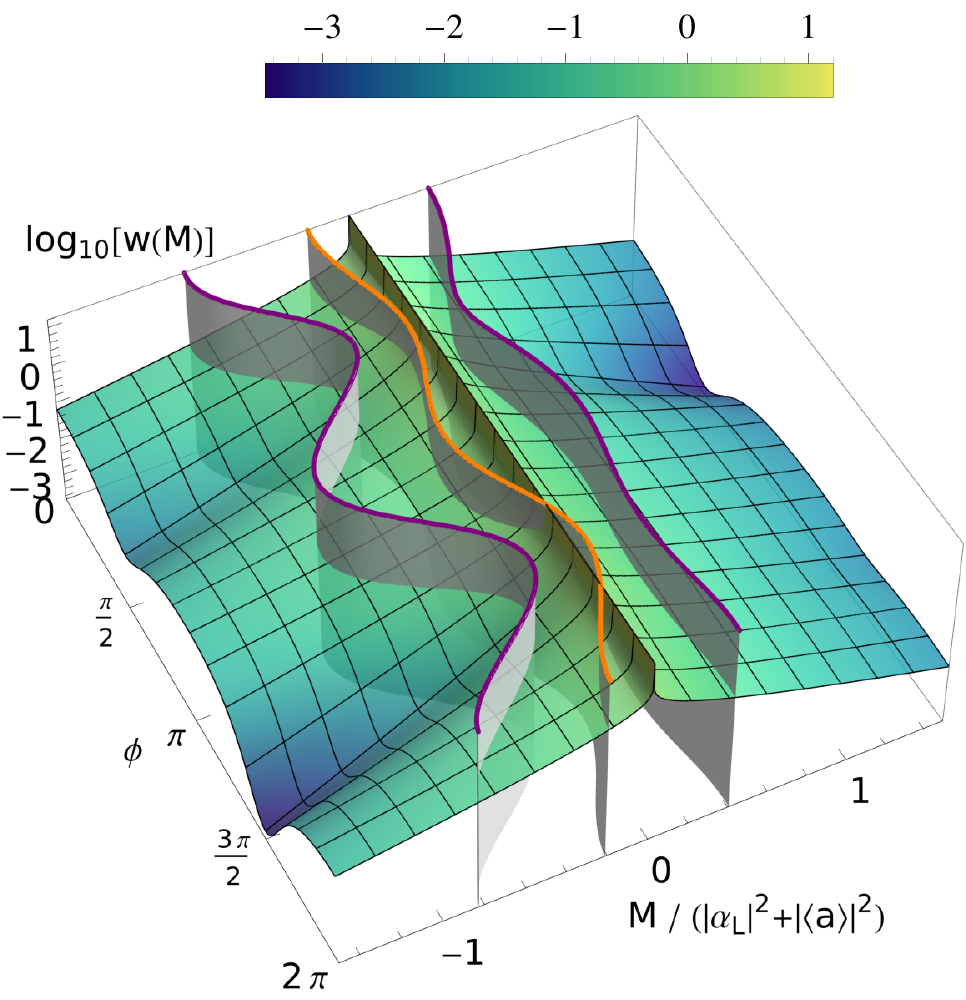}
			\label{fig:gaussAmplSq}
		\end{minipage}
 		\caption{
			Correlation statistics $w(M)$ as a function of the phase $\phi$ of the LO for a squeezed coherent state defined through Eq. \eqref{eq:characsq} with $V_{\mathrm{x}}=4.0$, $V_{\mathrm{p}}=0.5$, and mean amplitude $|\langle\hat a\rangle|$ equal to the amplitude $|\alpha_{\mathrm{L}}|$ of the strong LO, $|\alpha_{\mathrm{L}}|\gg 1$.
			Phase squeezing is shown on top ($\phi_\xi=\pi$) and amplitude squeezing on bottom ($\phi_\xi=0$).
			The plots are logarithmic (color-bar numbers indicate $\mathrm{log}_{10}[w(M)]$) and $M$ is normalized to $|\alpha_{\mathrm{L}}|^2+|\langle\hat a\rangle|^2$. 
			The orange line marks the expectation value $\mathbb{E}\left(M\right)$ and the purple lines indicate $\mathbb{E}\left(M\right)\pm\sqrt{\mathrm{var}\left(M\right)}$.
		}
		\label{fig:gauss}
	\end{figure}	
	
	\begin{figure}[t]
	\centering
 		\begin{minipage}[t]{1.0\linewidth}
			\includegraphics[width=0.99\linewidth]{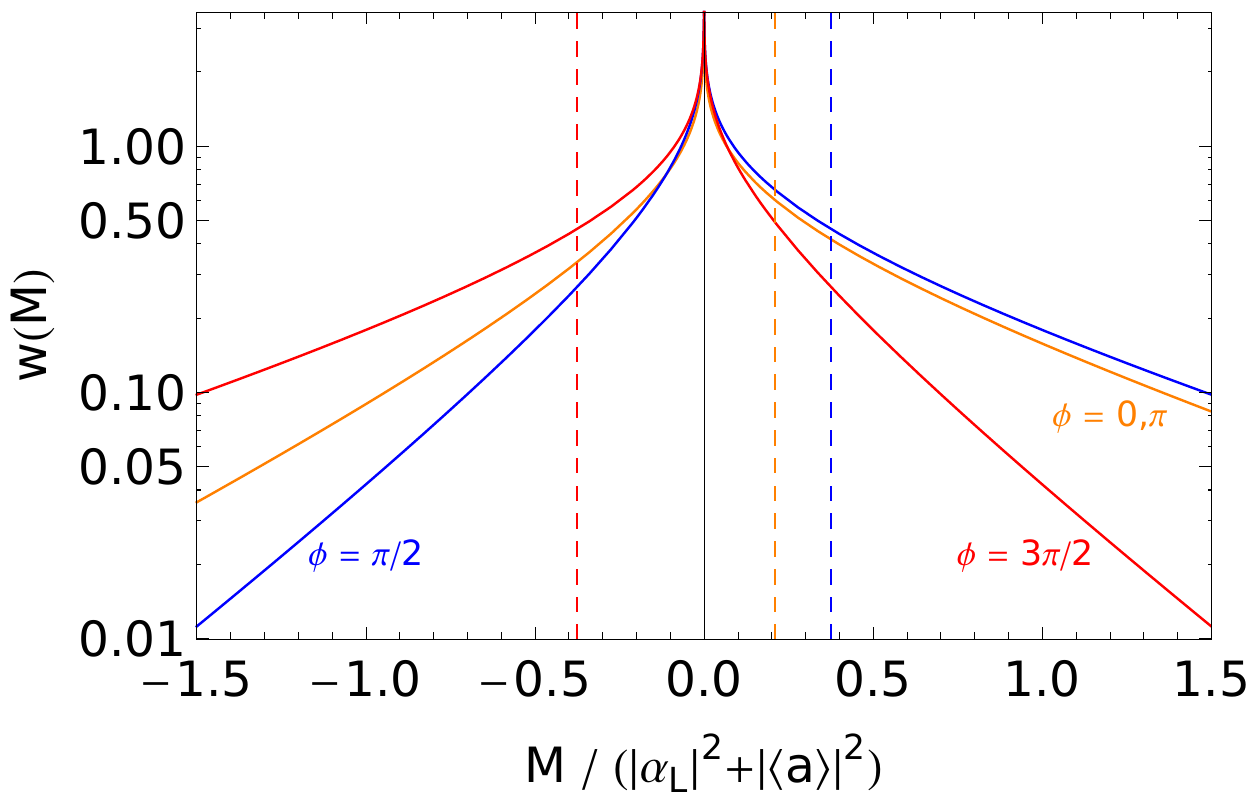}
			\label{fig:gaussphasePhaseSq}
 			\\\vspace*{3ex}
 			\includegraphics[width=0.99\linewidth]{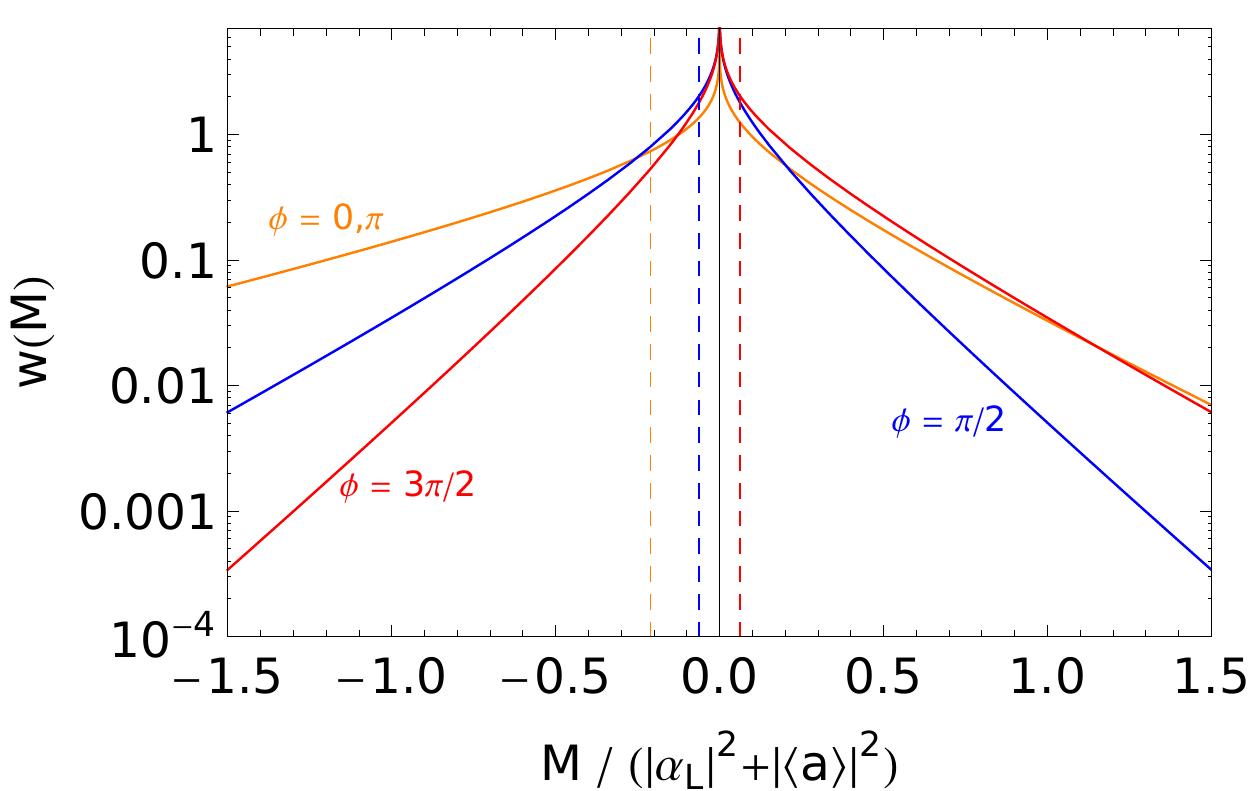}
 			\label{fig:gaussphaseAmplSq}
		\end{minipage}
 		\caption{
			Correlation statistics $w(M)$ as in Fig. \ref{fig:gauss} for various phases $\phi$ of the LO. 
			Phase squeezing is shown on top ($\phi_\xi=\pi$) and amplitude squeezing on bottom ($\phi_\xi=0$).
			The expectation values $\mathbb{E}(M)$ are indicated by the vertical dashed lines of the same color.
			Note that $M$ is normalized to $|\alpha_{\mathrm{L}}|^2+|\langle\hat a\rangle|^2$.
		}
		\label{fig:gaussphase}
	\end{figure}	
	
	This is a distribution of two correlated Gaussian variables with zero mean, the variances
	\begin{align}
		s_1^2&=J_{1,1},\\
		s_2^2&=J_{2,2}
	\end{align}
	and the correlation coefficient 
	\begin{align}
		\mathcal{C}&=-\dfrac{J_{1,2}}{\sqrt{J_{1,1}J_{2,2}}}.
	\end{align}
	Here the $J_{u,\ell}$ with $u,\ell=1,2$ are defined by
	\begin{align}
	\begin{aligned}
		J_{u,\ell}&=\sigma_u\sigma_\ell\delta_{u,\ell}\\
		&+\dfrac1{2}(-1)^{u+\ell}\left\{(V_{\mathrm{p}}+V_{\mathrm{x}}-2)\mathrm{Re}\left[h_u h_\ell^\ast\right]\right.\\
		&\left.+(V_{\mathrm{p}}-V_{\mathrm{x}})\mathrm{Re}\left[h_u h_\ell e^{-i\phi_{\xi}}\right]\right\}
	\end{aligned}
	\end{align}
	and they incorporate the variances $\sigma_j^2$ [cf. Eq. \eqref{eq:sigma}] and the quantities $h_j$ [cf. Eq. \eqref{eq:h}].
	Hence, we can directly apply the result of Ref. \cite{Cui2016} and find the closed expression for the correlation statistics
	\begin{align}
	\begin{aligned}
		w(M)&=\dfrac1{\pi s_1s_2\sqrt{1-\mathcal{C}^2}}\exp\left(\dfrac{\mathcal{C} M}{s_1s_2(1-\mathcal{C}^2)}\right)\\
		&\times K_0\left(\dfrac{|M|}{s_1s_2(1-\mathcal{C}^2)}\right).
	\end{aligned}
	\end{align}
	
	Figure \ref{fig:gauss} illustrates this probability distribution for the signal prepared in a phase- and amplitude-squeezed coherent state as a function of the phase of the LO for a realistic example.
	Here the homodyne cross-correlation scheme is applied, described by the matrix $\mathcal{Q}^{(\mathrm{cc})}$ in Eq. \eqref{eq:cc}.
	We set the beam-splitter transmittance-reflectance ratio to $|T|^2$:$|R|^2$=$14$:$86$, as it was applied in the experiment reported in Ref. \cite{Kuehn2017}.
	Furthermore, we assume ideal detectors, i.e., $\eta_1=\eta_2=1$ and $\nu_1=\nu_2=0$.
	The expectation value of $M$, which is given by
	\begin{align}
		\mathbb{E}\left(M\right)=\mathcal{C}s_1s_2,
	\end{align}
	and the standard deviation
	\begin{align}
		\sqrt{\mathrm{var}\left(M\right)}=\sqrt{1+\mathcal{C}^2}\,s_1s_2
	\end{align}
	are also shown in Fig. \ref{fig:gauss}.
	The correlation statistics for specific phases $\phi$ of the LO together with the expectation value $\mathbb{E}(M)$ is shown for both kinds of states in Fig. \ref{fig:gaussphase}.
	As for the coherent states, we obtain non-Gaussian correlation statistics for Gaussian signals.
			
\subsection{Fock states}
\label{sub:fock}

	\begin{figure}[h]
	\centering
 		\begin{minipage}[t]{1.0\linewidth}
			\includegraphics[width=0.99\linewidth]{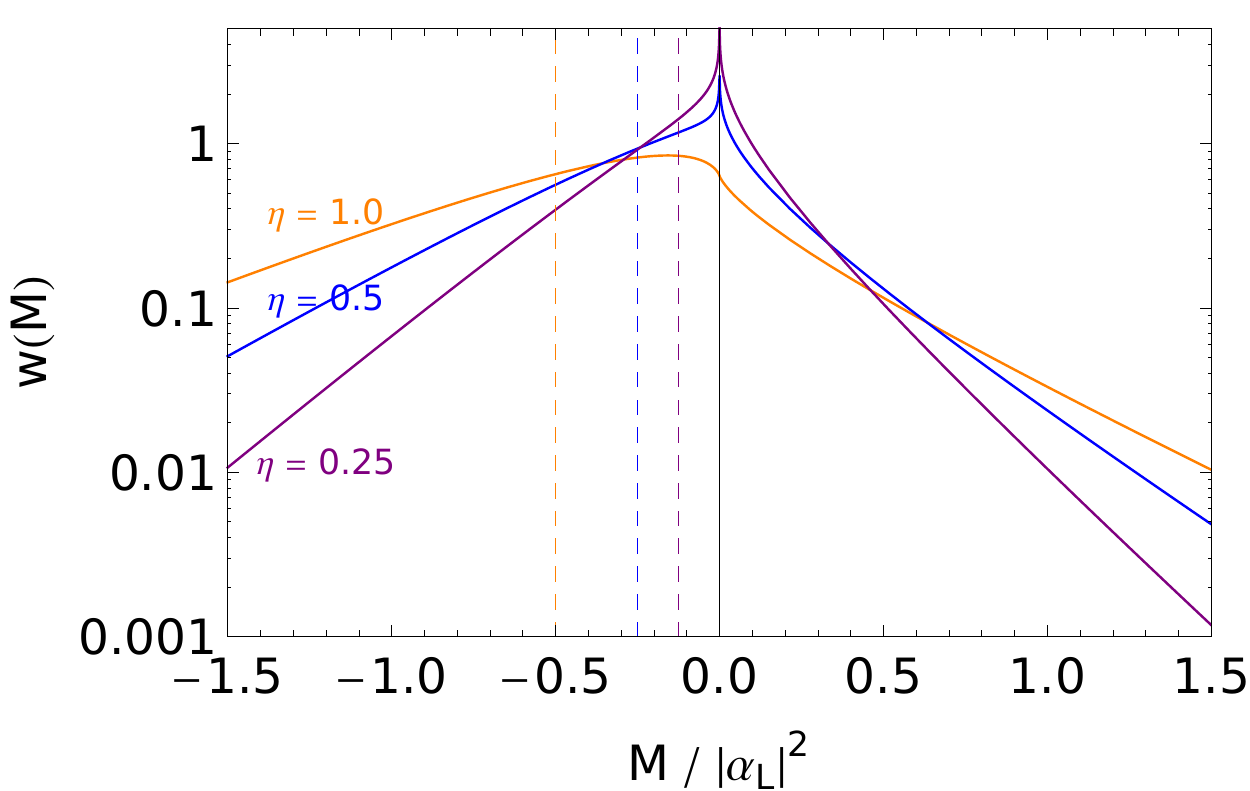}
			\label{fig:singlephoton}
 			\\\vspace*{3ex}
 			\includegraphics[width=0.99\linewidth]{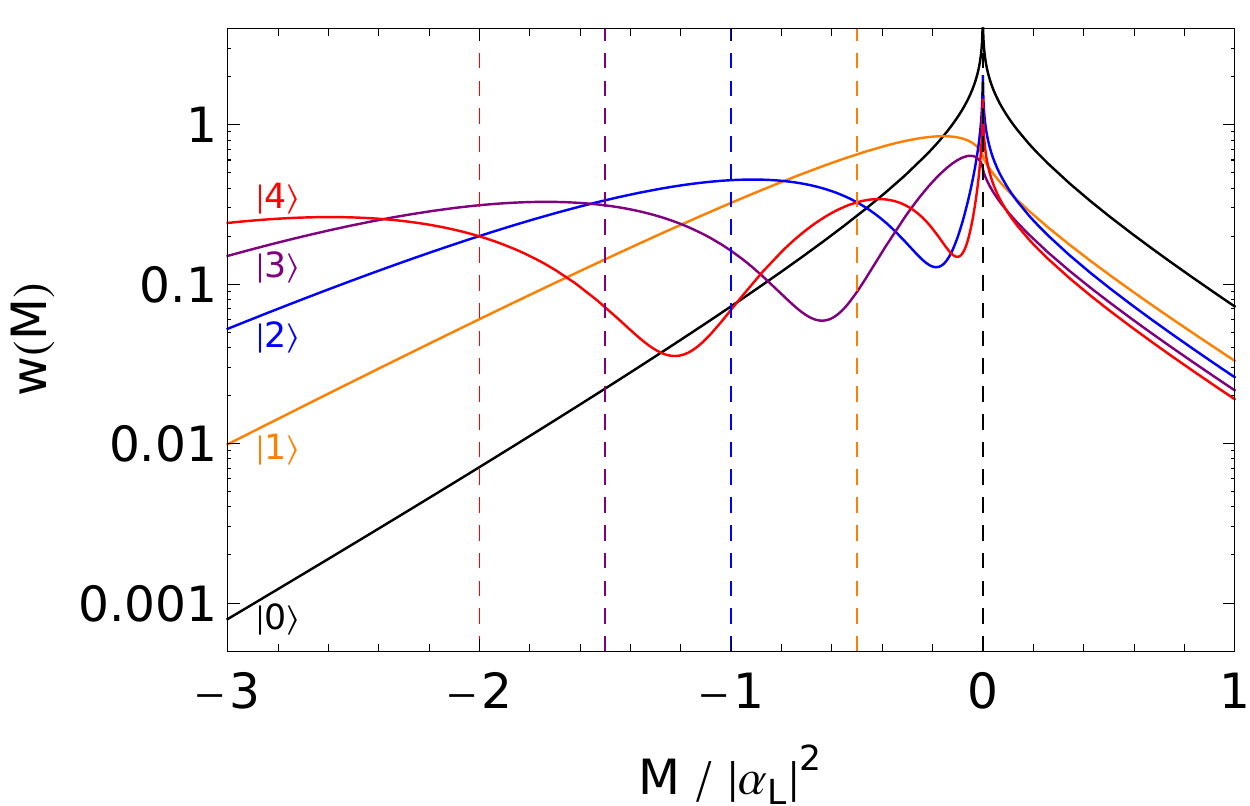}
 			\label{fig:multiphoton}
		\end{minipage}
 		\caption{
			Shown on top is the correlation statistics $w(M)$ for the signal beam prepared in a single-photon state.
			The two detectors in Fig. \ref{fig:Setup} have the same efficiencies $\eta_1=\eta_2$ and no dark noise counts.
			The statistics is shown for various total detector efficiencies, $\eta=\eta_1\cdot\eta_2$. 
			Shown on bottom is the correlation statistics for ideal detectors as a function of the number of photons $n$ in the signal beam for $n=0,1,2,3,4$.
			For both figures note that $M$ is normalized to the square of the absolute value of the LO amplitude.
			The expectation values $\mathbb{E}(M)$ are indicated by the vertical dashed lines of the same color.
		}
		\label{fig:photon}
	\end{figure}	

	The states, which are considered to be most contrary to the classical understanding of light, are the Fock states $|n\rangle$ ($n=1,2,\dots$), excluding the vacuum state.
	Let us calculate the resulting correlation statistics $w(M)$ if a Fock state impinges on our HCM detector in Fig. \ref{fig:Setup}. 
	Inserting the $P$~function of $|n\rangle$, which is given by
	\begin{align}\label{eq:FockP}
		P(\gamma)&=\sum_{q=0}^n\binom{n}{q}\dfrac1{q!}\partial_{\gamma}^q\partial_{\gamma^\ast}^q\delta(\gamma),
	\end{align}
	into Eq. \eqref{eq:fullwm}, one obtains, through integration by parts, the correlation statistics
	\begin{align}\label{eq:wmFock}
	\begin{aligned}
		w(M)&=\dfrac1{\sigma_1\sigma_2}\sum_{q=0}^n\binom{n}{q}\dfrac1{q!}\sum_{u=0}^q\sum_{\ell=0}^{2u}\mathcal{W}_{u,\ell}\left(\dfrac{M}{\sigma_1\sigma_2}\right)\\
		&\times\int d^2\gamma\,\delta(\gamma)\,\left[\partial_{\gamma}^q\partial_{\gamma^\ast}^q{\mathcal{G}}_{\ell,2u-\ell}(\gamma,\gamma^\ast)\right].
	\end{aligned}
	\end{align}
	The derivatives of $\mathcal{G}_{a,b}(\gamma,\gamma^\ast)$ with respect to $\gamma$ and $\gamma^\ast$, for $a,b=0,1,2,\dots$, are given by the formulas
	\begin{align}
	\begin{aligned}
		\partial_{\gamma^\ast}{\mathcal{G}}_{a,b}(\gamma,\gamma^\ast)&=\!-\!\left[\!\dfrac{h_1}{\sigma_1}\!\right]{\mathcal{G}}_{a+1,b}(\gamma,\gamma^\ast)\!-\!\left[\!\dfrac{h_2}{\sigma_2}\!\right]{\mathcal{G}}_{a,b+1}(\gamma,\gamma^\ast)\\
		&+\!a\!\left[\!\dfrac{h_1}{\sigma_1}\!\right]{\mathcal{G}}_{a-1,b}(\gamma,\gamma^\ast)\!+\!b\left[\!\dfrac{h_2}{\sigma_2}\!\right]{\mathcal{G}}_{a,b-1}(\gamma,\gamma^\ast)\\
		\partial_{\gamma}{\mathcal{G}}_{a,b}(\gamma,\gamma^\ast)&=\left[\partial_{\gamma^\ast}{\mathcal{G}}_{a,b}(\gamma,\gamma^\ast)\right]^\ast,
	\end{aligned}
	\end{align}
	with $\sigma_j$ and $h_j$ defined in Eqs. \eqref{eq:sigma} and \eqref{eq:h}, respectively.
	Together with
	\begin{align}
		\mathcal{G}_{a,b}(0,0)=\delta_{a,0}\delta_{b,0}
	\end{align}
	one can recursively evaluate the derivatives of $\mathcal{G}_{a,b}$ at $\gamma=0$ appearing in Eq. \eqref{eq:wmFock}.
	In particular, for a single photon the correlation statistics reads
	\begin{align}\label{eq:wmSinglePhoton}
	\begin{aligned}
		w(M)&=\dfrac1{\pi\sigma_1\sigma_2}\left\{\left[1-\dfrac{|h_1|^2}{\sigma_1^2}-\dfrac{|h_2|^2}{\sigma_2^2}\right]K_0\left(\dfrac{|M|}{\sigma_1\sigma_2}\right)\right.\\
		&+2\left[\dfrac{|h_1|^2}{\sigma_1^2}+\dfrac{|h_2|^2}{\sigma_2^2}\right]\dfrac{|M|}{\sigma_1\sigma_2}K_1\left(\dfrac{|M|}{\sigma_1\sigma_2}\right)\\
		&\left.+\dfrac1{\sigma_1\sigma_2}\left[h_1h_2^\ast+h_1^\ast h_2\right]\dfrac{|M|}{\sigma_1\sigma_2}K_0\left(\dfrac{|M|}{\sigma_1\sigma_2}\right)\right\}.
	\end{aligned}
	\end{align}
	Note that this general expression is employable in all types of homodyne correlation measurement schemes considered.
	A particular device can be specified by fixing the input-output matrix $\mathcal{Q}$ in Eq. \eqref{eq:inout}, which determines the values of $\sigma_j$ and $h_j$ in Eq. \eqref{eq:wmSinglePhoton} [cf. Eqs. \eqref{eq:sigma}, \eqref{eq:h}, and \eqref{eq:alphak}].
	
	Let us study in the following the particular scenario, where the signal Fock state is combined on a beam splitter of field strength transmittance $T$ and reflectance $R$ with a strong LO, $|\alpha_{\mathrm{L}}|\gg 1$.
	In this case the two detectors with efficiencies $\eta_1$ and $\eta_2$ and no dark noise counts receive high light intensities and the correlation statistics is independent of the phase of the LO.
	The expectation value of the measurement outcome $M$ as a function of the photon number $n$ can be calculated to be
	\begin{align}\label{eq:meanfock}
		\mathbb{E}(M)=-2\eta_1\eta_2|T|^2|R|^2|\alpha_\mathrm{L}|^2 n.
	\end{align}
	It decreases linearly with increasing photon number and is equal to zero for the vacuum state ($n=0$).
	
	Applying a $50$:$50$ beam splitter, the resulting correlation statistics $w(M)$ for the single photon ($n=1$) is shown for various detector efficiencies of the photodetectors in Fig. \ref{fig:photon} together with the mean value in Eq. \eqref{eq:meanfock}.
	We recognize the well-known linear dependence of $\mathbb{E}\left(M\right)$ on the detector efficiencies.
	By contrast, the shape of the statistics reveals a nonlinear dependence with respect to the efficiencies.
	In addition, the correlation statistics is shown in Fig.~\ref{fig:photon} for the signal beam prepared in various Fock states.
	We consider the same setup as previously with ideal detectors ($\eta_1=\eta_2=1$ and $\nu_1=\nu_2=0$).
	The mean values $\mathbb{E}\left(M\right)$ are nonpositive due to the fact that they are up to positive prefactors equal to the negative normal-ordered quadrature variance $-\langle:(\Delta \hat x)^2:\rangle$, which cannot be positive for Fock states (see also Ref. \cite{Vogel1995}).
		
\section{Nonclassicality tests}
\label{ch:nonclassicality}

	In this section we want to investigate whether there are nonclassical signatures in the correlation statistics $w(M)$.
	For this purpose, we first link our results of the previous sections to the certification of nonclassical effects via homodyne intensity-correlation and cross-correlation measurements.
	In this regard, we study in particular so-called anomalous quantum correlations.
	Afterward, we develop a criterion for the certification of nonclassical light based on the variance of the correlation statistics.

\subsection{Uncovering nonclassicality by the mean of the correlation statistics}
\label{sub:anomalous}

\begin{figure}[h]
	\centering
 		\includegraphics[width=0.99\linewidth]{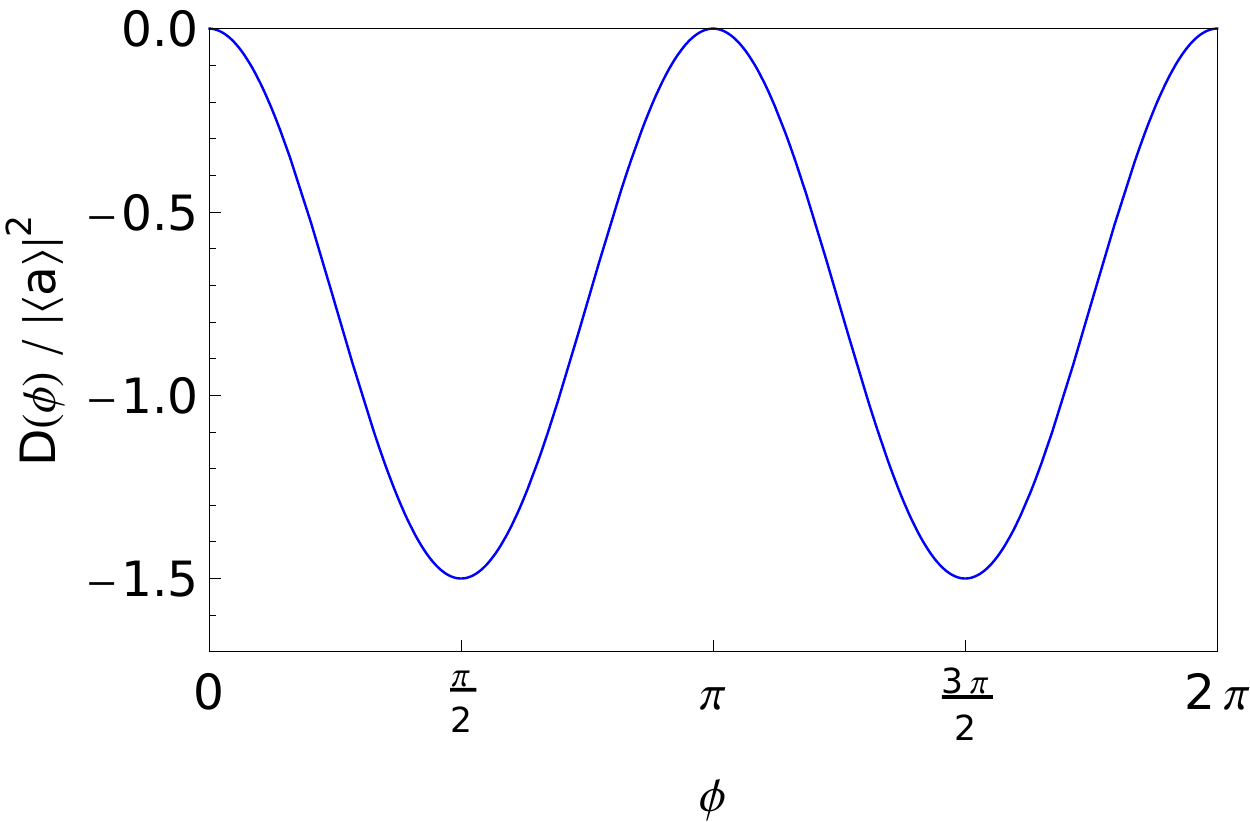}
 		\caption{
 			Value of $D(\phi)$ (normalized by $|\langle\hat a\rangle|^2$) in Eq. \eqref{eq:det} for a phase-squeezed coherent state as a function of the optical phase $\phi$.
			Negative values certify anomalous quantum correlations.
		}
		\label{fig:det}
 	\end{figure}

	Defining mixtures of coherent states $|\alpha\rangle$ as the classical reference, a state $\hat\rho$ is called nonclassical if its Glauber-Sudarshan $P$~function in the representation \eqref{eq:signal} does not have the properties of a probability density \cite{Titulaer1965}.
	It is already known that in the case of the homodyne intensity-correlation measurement associated with the matrix $\mathcal{Q}^{(\mathrm{ic})}$ in Eq. \eqref{eq:ic} of the LON, a negative expectation value $\mathbb{E}(M)$ of the correlation statistics directly indicates the nonclassicality of the signal field (for details see \cite{Vogel1995}).
	On the other hand, in the case of the homodyne cross-correlation measurement, described by the matrix $\mathcal{Q}^{(\mathrm{cc})}$ in Eq. \eqref{eq:cc} of the LON, the nonclassicality cannot be directly inferred from negativities of the mean correlation, $\mathbb{E}(M)$.
	However, one can decompose the latter in terms of various orders with respect to the LO field strength $\alpha_{\mathrm{L}}$, 
	\begin{align}\label{eq:meancorrhccm}
	\begin{aligned}
		\mathbb{E}(M)&=\eta_1\eta_2\left[|T|^2|R|^2\langle:(\Delta\hat n)^2:\rangle\right.\\
		&+|\alpha_{\mathrm{L}}||T||R|(|R|^2-|T|^2)\langle:\Delta\hat x_\phi\Delta\hat n:\rangle\\
		&\left.-|\alpha_{\mathrm{L}}|^2|T|^2|R|^2\langle:(\Delta\hat x_\phi)^2:\rangle\right],
	\end{aligned}
	\end{align} 
	with the optical phase $\phi$ as outlined in \cite{Vogel1995}. 
	Here $\langle:(\Delta\hat n)^2:\rangle$ and $\langle:(\Delta\hat x_\phi)^2:\rangle$ are the normal-ordered variances of the photon number $\hat n=\hat a^\dagger\hat a$ and the quadrature $\hat x_\phi=\hat a e^{i\phi}+\hat a^\dagger e^{-i\phi}$, respectively.
	The normal-ordered moment $\langle:\Delta\hat x_\phi\Delta\hat n:\rangle$ corresponds to the anomalous correlation of quadrature and photon-number fluctuations.
	One can distinguish two scenarios on the basis of the expression \eqref{eq:meancorrhccm}.
	If the LO intensity is much larger than the signal intensity, the expectation value $\mathbb{E}(M)$ corresponds to the negative normal-ordered quadrature variance $-\langle:(\Delta\hat x_\phi)^2:\rangle$ of the signal, which is why a positive expectation value of $M$ indicates squeezing.
	By contrast, if the LO intensity is comparable to the intensity of the signal beam also the normal-ordered moments $\langle:(\Delta\hat n)^2:\rangle$ and $\langle:\Delta\hat x_\phi\Delta\hat n:\rangle$ contribute in Eq. \eqref{eq:meancorrhccm}.
	Methods to extract the three moments $\langle:(\Delta\hat n)^2:\rangle$, $\langle:\Delta\hat x_\phi\Delta\hat n:\rangle$, and $\langle:(\Delta\hat x_\phi)^2:\rangle$ from the mean \eqref{eq:meancorrhccm} of the correlation statistics have been proposed in \cite{Vogel1995}.
	These techniques have recently been successfully applied in an experiment \cite{Kuehn2017} and the separated moments have been used to test the violation of the Cauchy-Schwarz inequality 
	\begin{align}\label{eq:det}
		D(\phi)=\langle:(\Delta\hat n)^2:\rangle\langle:(\Delta\hat x_\phi)^2:\rangle-\langle:\Delta\hat x_\phi\Delta\hat n:\rangle^2\stackrel{\mathrm{cl}}{\geq} 0,
	\end{align}
	which is fulfilled for all classical states.
	As this condition incorporates an anomalous moment of two noncommuting observables, a violation of the inequality refers to the presence of anomalous quantum correlations.
	Remarkably, anomalous quantum correlations of a phase-squeezed coherent state have been certified experimentally, for almost the full range of the optical phase $\phi$, i.e., even for phases corresponding to antisqueezing \cite{Kuehn2017}. 
	This is consistent with the theoretical prediction.
	The three moments under consideration are given for a general Gaussian state, as defined by Eq. \eqref{eq:characsq}, with large mean value ($|\langle\hat a\rangle|\gg 1$) by
	\begin{align}
		\langle:(\Delta\hat n)^2:\rangle&=|\langle\hat a\rangle|^2\left[\dfrac{V_{\mathrm{p}}-V_{\mathrm{x}}}{2}\cos[2\arg(\langle\hat a\rangle)-\phi_\xi]\right.\notag\\
		&\left.+\dfrac{V_{\mathrm{x}}+V_{\mathrm{p}}-2}{2}\right],\\
		\langle:\Delta\hat x_\phi\Delta\hat n:\rangle&=|\langle\hat a\rangle|\left[\dfrac{V_{\mathrm{p}}-V_{\mathrm{x}}}{2}\cos[\phi-\arg(\langle\hat a\rangle)+\phi_\xi]\right.\notag\\
		&\left.+\dfrac{V_{\mathrm{x}}+V_{\mathrm{p}}-2}{2}\cos[\phi+\arg(\langle\hat a\rangle)]\right],\\
		\langle:(\Delta\hat x_\phi)^2:\rangle&=\dfrac{V_{\mathrm{p}}-V_{\mathrm{x}}}{2}\cos(2\phi+\phi_\xi)+\dfrac{V_{\mathrm{x}}+V_{\mathrm{p}}-2}{2}.
	\end{align}
	Figure \ref{fig:det} shows the resulting quantity $D(\phi)$ in \eqref{eq:det} for a phase-squeezed coherent state ($\phi_\xi=\pi$) with the orthogonal quadrature variances $V_{\mathrm{x}}=4.0$ and $V_{\mathrm{p}}=0.5$, which are the parameters also used in Sec. \ref{sub:gaussian}.
	The negativity of $D(\phi)$, except for phases $\phi$ being multiples of $\pi$, unambiguously uncovers anomalous quantum correlations.
		
\subsection{Higher order quantum features of the correlation statistics}
\label{sub:ncl}
		
	Our knowledge of the full statistics of $M$ allows us to use higher moments beyond the expectation value to visualize nonclassical effects.
	Suppose the state of the signal field is classical, i.e., its $P$~function is a classical probability distribution $P_{\mathrm{cl}}(\gamma)$. 
	Recalling Eqs. \eqref{eq:expcond}--\eqref{eq:Mvar}, we can use the relation
	\begin{align}
	\begin{aligned}
		\mathrm{var}\left(M\right)&=\int d^2\gamma\,P_{\mathrm{cl}}(\gamma+\langle\hat a\rangle)\\
		&\times\left\{\left[\mathbb{E}\left(M|\gamma\right)-\mathbb{E}\left(M\right)\right]^2+\mathrm{var}\left(M|\gamma\right)\right\}
	\end{aligned}
	\end{align}
	for mixture distributions, where $\mathbb{E}\left(M|\gamma\right)$ and $\mathrm{var}\left(M|\gamma\right)$ are the expectation value and the variance of $M$ conditioned on the value of $\gamma$, respectively.
	Obviously,
	\begin{align}\label{eq:varMest}
	\begin{aligned}
		\mathrm{var}\left(M\right)&\geq \int d^2\gamma\,P_{\mathrm{cl}}(\gamma+\langle\hat a\rangle)\,\mathrm{var}\left(M|\gamma\right)\\
		&\geq \mathrm{min}_{\gamma}\,\mathrm{var}\left(M|\gamma\right)
	\end{aligned}
	\end{align}
	holds for all classical states.
	The conditional variance is derived from Eqs. \eqref{eq:expcond} and \eqref{eq:secondcond} as
	\begin{align}
	\begin{aligned}
		\mathrm{var}\left(M|\gamma\right)&=\mathbb{E}\left(M^2|\gamma\right)-\left[\mathbb{E}\left(M|\gamma\right)\right]^2\\
		&=\sigma_1^2\sigma_2^2\left[1+\left(\dfrac{\mu_1(\gamma)}{\sigma_1}\right)^2+\left(\dfrac{\mu_2(\gamma)}{\sigma_2}\right)^2\right].
	\end{aligned}
	\end{align}	
	It is minimal for $\gamma=0$, resulting in $\mu_1=\mu_2=0$ [cf. Eq.~\eqref{eq:mean}], and we arrive at
	\begin{align}\label{eq:minvar}
		\mathrm{min}_{\gamma}\,\mathrm{var}\left(M|\gamma\right)&=\sigma_1^2\sigma_2^2,
	\end{align}
	which is the same expression one obtains if the signal is in a coherent state with amplitude $\langle\hat a\rangle$.
	Introducing the quantity
	\begin{align}\label{eq:r}
		r=\dfrac{\mathrm{var}\left(M\right)}{\sigma_1^2\sigma_2^2}-1,
	\end{align}
	we infer from \eqref{eq:varMest} together with Eq. \eqref{eq:minvar} that all classical states fulfill the inequality
	\begin{align}\label{eq:rcl}
		r\stackrel{\mathrm{cl}}{\geq}0.
	\end{align}
	By contrast, if
	\begin{align}\label{eq:ncltest}
		r<0
	\end{align}
	is observed in the experiment, nonclassicality of the state under study is certified.
	Note that this requires an additional measurement with the signal beam prepared in a coherent state $|\alpha\rangle$ with its amplitude equal to the mean amplitude of the state under study, i.e., $\alpha=\langle\hat a\rangle$.
	In this measurement the same amplitude of the LO is used as for the measurement with the signal state under study.
	
	\begin{figure}[h]
	\centering
 		\includegraphics[width=0.99\linewidth]{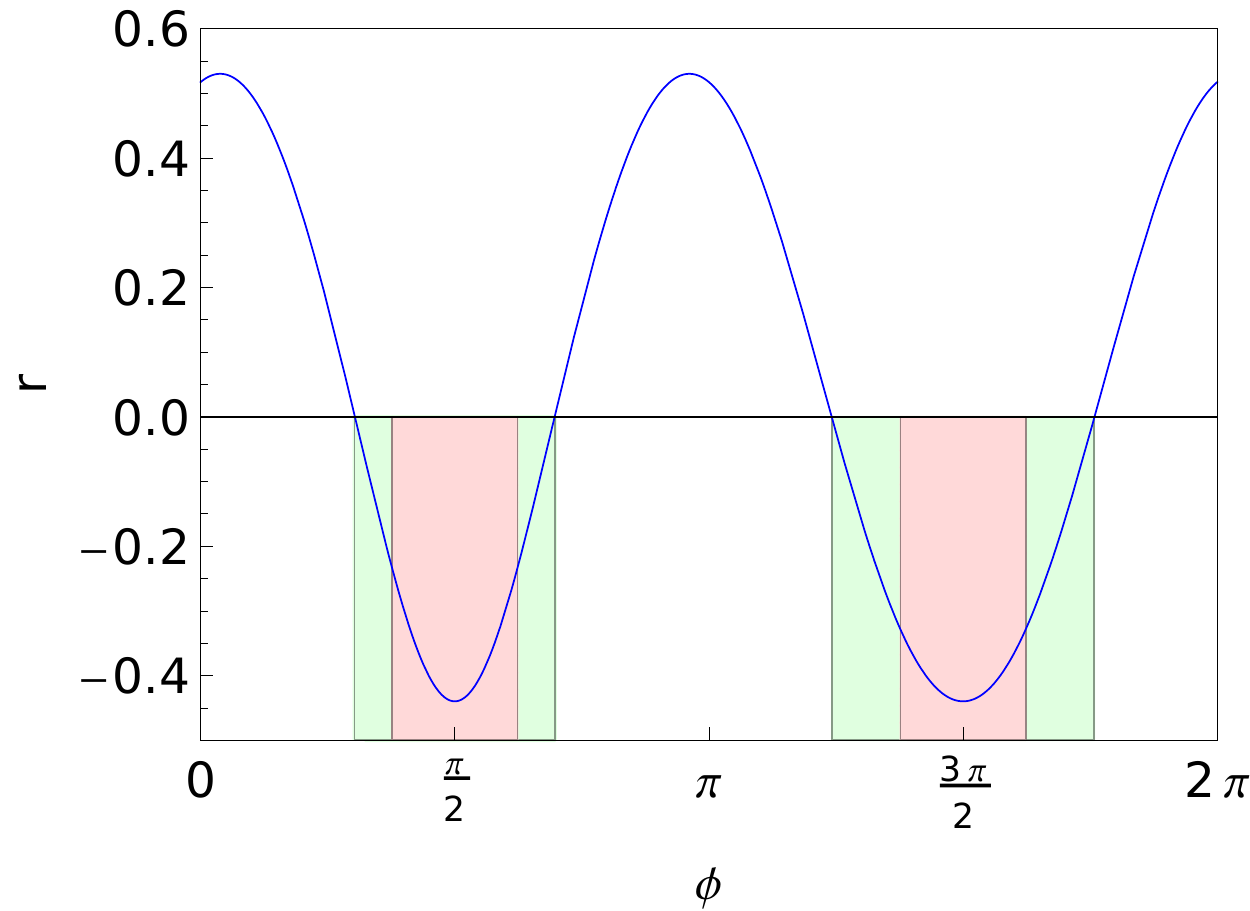}
		\caption{
			Quantity $r$, defined in Eq. \eqref{eq:r} as a function of the LO phase $\phi$ for an amplitude-squeezed coherent state, probed by a homodyne cross-correlation measurement.
			Squeezing is present for the phases in the light red colored region.
			The criterion in Eq. \eqref{eq:ncltest} reveals nonclassicality not only in the light red colored, but also in the extended light green colored phase region.
		}
		\label{fig:ncl}
	\end{figure}
	
	In the preceding section, we studied the certification of the nonclassicality of the signal field for the particular case of the homodyne cross-correlation measurement with weak LO, i.e., the signal and the LO have comparable intensities.
	These considerations focus on the mean value of the correlation statistics.
	Since this mean value does not directly uncover nonclassical effects, three different normal-ordered moments of the signal quadrature and photon number are separated from this quantity to show the violation of the classicality condition~\eqref{eq:det}.
	Note that the extraction of these moments requires precise knowledge of the beam-splitter transmittance-reflectance ratio and also measurements for different LO amplitudes or phases (see Ref.~\cite{Kuehn2017} for details).
	Such a separation procedure is not necessary if the variance of the correlation statistics is used to show the nonclassicality via the condition~\eqref{eq:ncltest}.
	Figure~\ref{fig:ncl} shows the value of $r$ as a function of the LO phase for the amplitude-squeezed coherent state defined by Eq. \eqref{eq:characsq} with $V_{\mathrm{x}}=4.0$, $V_{\mathrm{p}}=0.5$, and $\phi_\xi=0$, which was considered also in Sec. \ref{sub:gaussian}.
	This result corresponds to the detection via homodyne cross-correlation measurement with the mean amplitude $|\langle\hat a\rangle|$ of the signal being equal to the strong LO amplitude $|\alpha_{\mathrm{L}}|\gg 1$, with ideal photodetectors and with the beam-splitter transmittance-reflectance ratio set to $|T|^2$:$|R|^2$=$14$:$86$.
	We observe that $r$ is negative for a wider region of LO phases $\phi$ than the phase region where squeezing is present.
	This demonstrates the nonclassicality of this state by our criterion in Eq.~\eqref{eq:ncltest} for an extended range of LO phases, beyond the range of squeezing.
	For comparison, the nonclassicality test based on the violation of condition~\eqref{eq:det} is for this state even more powerful as the nonclassicality is shown for almost all optical phases (see Fig.~\ref{fig:det}). 
	However, the nonclassicality condition~\eqref{eq:ncltest} is directly based on the variance of the correlation statistics and thus does not require additional measurements with other LO configurations to extract further information, as needed to test condition~\eqref{eq:det}.
	It is expected that nonclassicality criteria involving higher moments, $\mathbb{E}\left(M^k\right)$ with $k>2$, will uncover nonclassicality of a larger class of states. 

\section{Conclusions}\label{ch:Conclusions}

	Motivated by the successful implementation of homodyne correlation measurements, we provided a rigorous derivation of the full statistics for the outcome of such a class of measurement devices, given high light intensities incident on the employed photodetectors.
	This correlation statistics is associated with a non-Gaussian POVM, which we determined in this work explicitly.
	In this regard, the probability distribution is non-Gaussian if the correlation detector is fed with Gaussian states of light, such as coherent states.
	Additionally, the results for squeezed coherent states and Fock states in the signal beam were calculated.
	We retrieved the linear dependence of the expectation value of the correlation statistics on the quantum efficiency, but showed also that the shape of the whole statistics strongly depends on the amount of loss. 
	
	Extending nonclassicality tests based on the mean of the correlation statistics, we formulated a nonclassicality condition, which is based on the variance of this statistics.
	We demonstrated the usefulness of this higher-order condition to certify the nonclassicality of an amplitude-squeezed coherent state.
	It is an open and interesting matter whether the correlation statistics for all LO phases contains the whole quantum information of the probed field, as is the case for the difference statistics in balanced homodyne detection.
	This and related questions should be addressed by further research.
	
\begin{acknowledgements}
	This work has received funding from the European Union's Horizon 2020 research and innovation program under Grant Agreement No. 665148 (QCUMbER).
\end{acknowledgements}

\end{document}